\documentclass[a4paper,twoside]{article}

\usepackage{alltt}
\usepackage{newfloat}
\usepackage{caption}

\DeclareFloatingEnvironment[fileext=frm,placement={!ht},name=Listing]{example}
\DeclareCaptionSubType*{example}
\usepackage{epsfig}
\usepackage{subcaption}
\usepackage{calc}
\usepackage{amssymb}
\usepackage{amstext}
\usepackage{amsmath}
\usepackage{amsthm}
\usepackage{multicol}
\usepackage{pslatex}
\usepackage{apalike}
\usepackage{multicol, blindtext}
\usepackage{SCITEPRESS}     % Please add other packages that you may need BEFORE the SCITEPRESS.sty package.

\newtheoremstyle{mytheoremstyle} 
    {\topsep}                    
    {\topsep}                    
    {\fontfamily{tm}}                   
    {}                           
    {\scshape\bfseries}                   
    {.}                          
    {.5em}                       
    {}  

\theoremstyle{mytheoremstyle}
\newtheorem{thm}{Theorem}

\theoremstyle{mytheoremstyle}
\newtheorem{defin}{Definition}

\begin{document}

\title{Computer Viruses: The Abstract Theory Revisited}

\author{\authorname{Nikolai Gladychev\orcidAuthor{0000-0002-9744-6169}}
\affiliation{Department of Computer Science, University College Dublin, Belfield, Dublin 4, Ireland}
\email{nikolai.gladychev@gmail.com}
}


\abstract{Identifying new viral threats, and developing long term defences against current and future computer viruses, requires an understanding of their behaviour, structure and capabilities. This paper aims to advance this understanding by further developing the abstract theory of computer viruses. A method of providing abstract definitions for classes of viruses is presented in this paper, which addresses inadequacies of previous techniques. Formal definitions for some classes of viruses are then provided, which correspond to existing informal definitions. To relate the abstract theory to the real world, the connection between the abstract definitions and concrete virus implementations is examined. The use of the proposed method in studying the fundamental properties of computer viruses is discussed.}

\onecolumn \maketitle \normalsize \setcounter{footnote}{0} \vfill

\section{\uppercase{Introduction}}
\label{sec:1}
\noindent Current antiviral detection methods and techniques are largely reactive, with antivirus software being updated according to new viruses and threats that are discovered\cite{FiliolText}\cite{Dechaux}. There is an ``arms race'' between computer virus and antivirus writers\cite{Kramer}, and any antiviral techniques developed for current computer virus, are ultimately bypassed by new, and more advanced viral behaviours. A more proactive approach would be to detect new threats before they emerge in the real world, for which a thorough understanding of the possible behaviours, structures and capabilities of computer viruses is required. It is the aim of the abstract theory of computer viruses to view the underlying mechanisms and principles of computer viruses independently of implementation complexities, and to provide some more general results about computer viruses. And it is the aim of this paper to further this abstract understanding of computer viruses.   

Malware is a more general concept than viruses, and computer viruses are commonly understood as malware which has some kind of self-replicating or self-propagating mechanism\cite{FiliolText}\cite{Cohen}. Nevertheless computer viruses remain highly relevant to the modern context, with network propagating trojans(worms), and botnets relating to viruses. Computer virus models can be extended to capture malware in general by allowing for non-replicating programs\cite{Adleman}, however this paper is concerned with the self-replicating case. The main contributions of this paper are as follows.
\begin{itemize}
\item A formal method for specifying computer viruses which has its roots in computability theory is proposed. The possible specifications are broader in scope and are more expressive than those possible using previous methods.
\item A technique for the construction of virus implementations from the formal descriptions is demonstrated.
\end{itemize}

Section \ref{sec:2} of the paper reviews how recursion theory(also known as computability theory) relates to self-replicating programs and computer viruses. Section \ref{sec:3-2} presents a formal framework and methodology for specifying computer viruses. It is the result of addressing some previously unnoted inadequacies of related work, which are discussed in section \ref{sec:3-1}. Section \ref{sec:4} goes on to use this framework to provide formal counterparts to a number of informal classifications of computer viruses. In particular, classes which could not be formally specified using previous methods are presented. Section \ref{sec:4-2} demonstrates how the framework in this paper can be used to study fundamental aspects of structure and behaviour of computer viruses. More complicated viruses require more complicated theorems from the recursion theory, and these theorems are discussed in section \ref{sec:5}. In section \ref{sec:5-example}, the implementation of a virus from a relatively complicated abstract definition is constructed, thereby illuminating the connection between the abstract theory and the real world. Section \ref{sec:6} provides concluding remarks.  

\section{\uppercase{Computer Viruses in Recursion Theory}}
\label{sec:2}
\noindent This paper will describe computer viruses in functional terms using standard mathematical notation. In the real world, like any program, computer viruses appear as some sequence of instructions. Partial recursive functions are those functions which can be computed by some sequence of instructions\footnote{More correctly: any function that can be computed using some system of Turing complete data-manipulation rules.}\cite{Rogers87}, and the theory of recursive functions allows for some manipulation of these sequences. Of particular interest in this paper, is how Kleene's second recursion theorem can be used to produce viruses. This approach has been used before\cite{Zuo&Zhuo}\cite{Bonfante}, however while constructive proofs for the existence of certain viruses have been produced, generation of concrete programs from these proofs is not straightforward\cite{Kraus}. To partially bridge the gap between this abstract approach and the realities of implementation, this section provides an intuitive explanation of the proof of Kleene's second recursion theorem, and outlines how programs can be produced from this construction. A more thorough treatment of generation of viruses is given in section \ref{sec:5} for a more complicated recursion theorem. 

In formal models of computer viruses thus far, a defining feature of a computer virus is the ability for self-replication\cite{Cohen}\cite{Adleman}. As a trivial example of a program with a self-replicating element consider a program which outputs its own sequence of instructions\footnote{This kind of program is known as a ``quine''.}. To construct it, the partial recursive function $f$ which takes two arguments and outputs its first argument(i.e. $f(x, y) = x$) can be used.

In pseudocode the instructions for $f$ could be:

\begin{small}
\begin{verbatim}
f(x, y)
 1:  Begin
 2:  return x
 3:  End
\end{verbatim}
\end{small}
\noindent Any input can be given as $x$, including the sequence of instructions for $f$. 
\\\indent Consider now a function which is the same as $f$, except that it has its own sequence of instructions ``hardcoded'' into its sequence of instructions so that it only takes the one argument $y$. Let $\textbf{e}$ denote the sequence of instructions for this function, and let $\varphi_\textbf{e}$ denote the function computed by $\textbf{e}$. The naive approach of ``hardcoding`` is quite troublesome\cite{Kraus}.

This approach has the structure:

\begin{small}
\begin{alltt}
\(\varphi\sb{\mathbf{e}}(y)\)
 1:  Begin
 2:  x \(\leftarrow\) sequence of instructions \(\mathbf{e}\) 
 3:  return x
 4:  End
\end{alltt}
\end{small}

Which expands infinitely into

\begin{small}
\begin{alltt}
\(\varphi\sb{\mathbf{e}}(y)\)
 1:  Begin
 2:  x \(\leftarrow\) ``Begin; x \(\leftarrow\) 
             ``Begin; x \(\leftarrow\)
             ``Begin; x \(\leftarrow\)
             ...''
 3:  return x
 4:  End
\end{alltt}
\end{small}
The solution instead lies in including an algorithm within the instructions which performs the hardcoding itself.

This approach has the structure:

\begin{small}
\begin{alltt}
\(\varphi\sb{\mathbf{e}}(y)\)
 1:  Begin
 2:  x \(\leftarrow\) (sequence of instructions \(\mathbf{e}\)
             with line 2 omitted)
 3:  out \(\leftarrow\) (everything up to line 2)
 4:  out \(\leftarrow\) out + ``x \(\leftarrow\) ''
 5:  out \(\leftarrow\) out + x
 6:  out \(\leftarrow\) out + (rest of the instructions
                     starting at line 3)
 7:  return out  
 8:  End
\end{alltt}
\end{small}
This solution for the construction of $\textbf{e}$ explains the essence of Kleene's recursion theorem, when it is observed that the outputs $f(\textbf{e}, y)$ and $\varphi_\textbf{e}(y)$ are the same thing: the sequence of instructions $\textbf{e}$. Whereas a specific $f$ was given above, Kleene's theorem captures the general case.

\begin{figure}[!h]
  \centering
   {\epsfig{file = 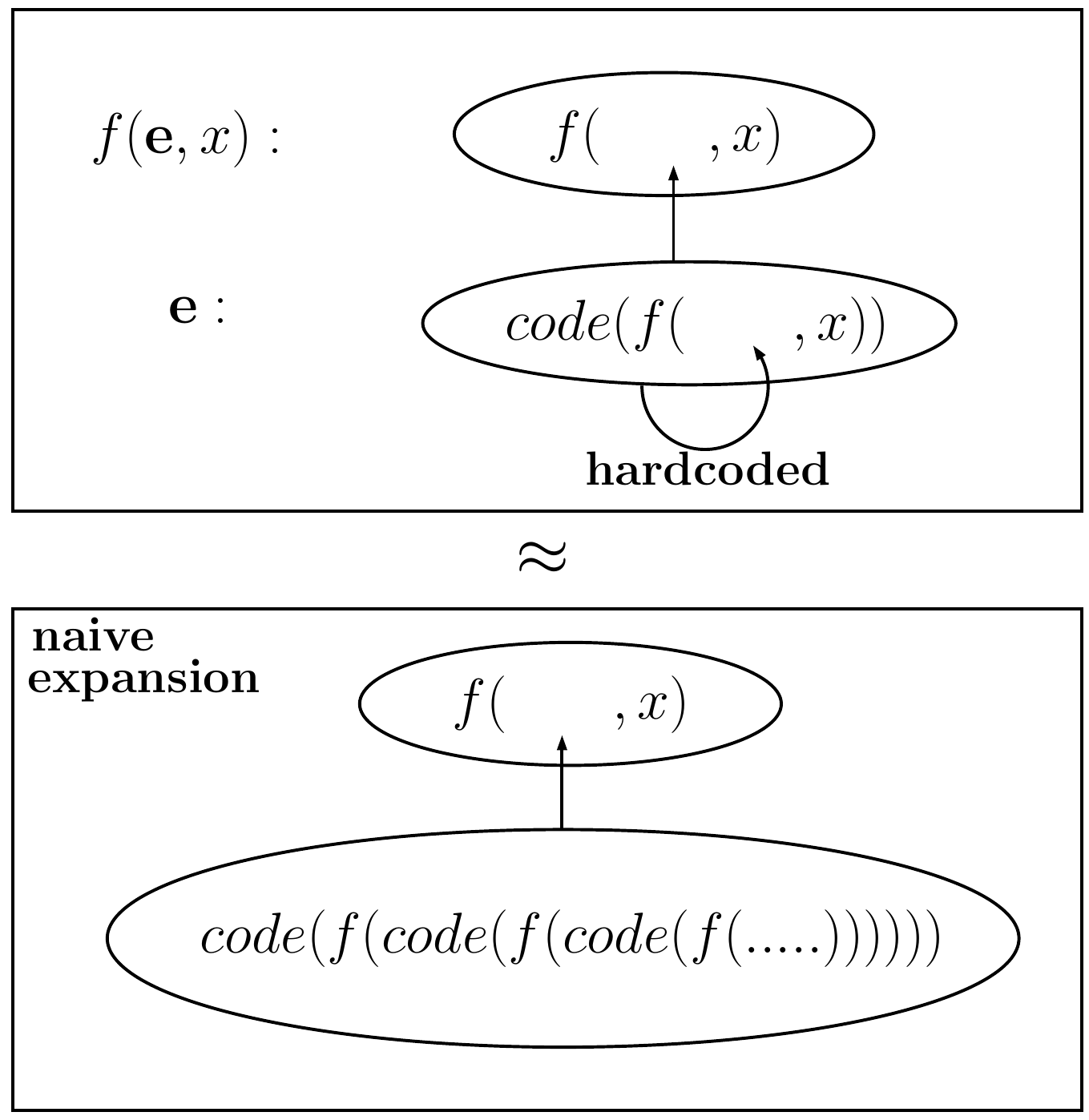, width = 5.5cm}}
  \caption{Depiction of the construction for Theorem \ref{th:Kleene}.}
  \label{fig1}
 \end{figure}

\begin{thm}[\bfseries Kleene's 2$^{nd}$ Recursion Theorem]\label{th:Kleene}
If $f$ is a partial recursive function, then there is a sequence of instructions {\bf{e}} such that\\ $\varphi_{\bf{e}}(x) = f({\bf{e}}, x)$.
\end{thm}
\noindent Kleene's theorem states that an $\textbf{e}$ can be found for any $f$. The method for constructing $\textbf{e}$ will be similar to the method shown above for the specific instance of $f$. It consists of finding the sequence of instructions for a function similar to $f$ except that it has a hardcoded value instead of its first argument(hence the function takes one fewer arguments), where the hardcoded value is that same sequence of instructions. An algorithmic solution is required, whereby the hardcoding process is included in the sequence of instructions(as shown above). A graphical interpretation of the construction for the proof of Kleene's recursion theorem appears in Figure \ref{fig1}, where the informal notation $code(f)$ is used to denote the sequence of instructions for $f$, so that for all $x$ and $y$, $\varphi_{code(f)}(x, y) = f(x,y)$. 

To see how this relates to computer viruses, define a rudimentary computer system environment as a tuple consisting of some number of data files and some number of program files: $(d_1, ..., d_n, p_1, ..., p_m)$. Then define a program within a system environment as a sequence of instructions which compute a function which takes that system environment, and outputs that same environment with some possible modifications. An example of a file overwriting virus, would be a sequence of instructions which compute a function that takes a system environment, and returns that system environment with all the program files replaced with the virus\footnote{In this case the infected form is exactly the viral sequence of instructions. The host program is completely overwritten.}. Kleene's theorem constructs this virus as follows: take the function $f$ defined as \[f(x , d_1, ..., d_n, p_1, ..., p_m) = (d_1, ..., d_n, x, ..., x).\]\noindent Apply the theorem to obtain the virus $\textbf{e}$, since $\textbf{e}$ as defined in the theorem satisfies: \[\varphi_{\textbf{e}}(d_1, ..., d_n, p_1, ..., p_m) = (d_1, ..., d_n, \textbf{e}, ..., \textbf{e}).\]

The theorem can prove the existence of viruses for any Turing complete system, and does not assume the existence of an operating system, or the ability to read or write files. In practice, applying the theorem to real computer programs is simpler, since a program may simply read its own sequence of instructions from file instead of ``hardcoding'' them. However Theorem \ref{th:Kleene} guarantees that this mechanism is not absolutely necessary.

\section{\uppercase{Abstract Description of Computer Viruses}}
\label{sec:3}
\noindent This section presents a novel way of describing the behaviour of various viruses in terms of partial recursive functions, such that real computer viruses can be constructed from these descriptions with Theorem \ref{th:Kleene}(as discussed in the previous section). The abstraction in this approach allows for the study of defining traits which classify types of viruses independently of their implementation.

\subsection{Related Work}
\label{sec:3-1}

The abstract theory of computer viruses was established by Cohen in \cite{Cohen} and Adleman in \cite{Adleman}. Cohen used a Turing Machine formalism, and loosely\footnote{Cohen also provides a formal definition, which is considerably more involved than its ``loose'' counterpart. However the same essential idea is captured.} defined a ``virus'' as a sequence of symbols which when interpreted in a given environment causes another sequence of symbols to be modified to contain a (possibly evolved) form of the virus. This definition is very general and implicitly captures viruses for any mode of infection. Following Cohen's work and with reference to specific viral behaviours, Adleman used partial recursive functions to provide a definition for computer viruses. That method was in turn extended by Zuo and Zhuo in \cite{Zuo&Zhuo}, where more specific objects describing aspects of computer viruses were introduced, which allowed for the formal definition of some classes of viruses. Adleman and Zuo and Zhuo viewed viruses as mappings from programs into ``infected programs'', and did not consider viruses independently of a host program. Viral programs appear independently of a host in another recursion theoretic approach in \cite{Bonfante}, in which an alternative to Adleman's definition is provided as it was found to be too restrictive.

There is a concept in \cite{Bonfante} that is thought of as an infected form, called the ``propagation vector'', denoted $\mathcal{B}(v, p)$. Viruses are defined with respect to a propagation vector, which describes how a virus infects a program. However while $\mathcal{B}(v, p)$ is viewed as the ``infected form'' of the program $p$ by the virus $v$ within the formalism, it is argued in this paper that it does not adequately correspond to the informal notion and practical reality of an ``infected form''. To demonstrate this, the definition for virus in \cite{Bonfante} is here reproduced:\\

\begin{defin}[\bfseries Virus w.r.t Propagation Vector]\label{def:Bonfante}
Assume that $\mathcal{B}$ is a partial recursive function. A virus w.r.t. to $\mathcal{B}$ is a program $\mathbf{v}$ such that for any tuple of programs $\mathbf{p}$, $x_1$, ..., $x_n$,
\[
\varphi_{\mathbf{v}}(\mathbf{p}, x_1, ..., x_n) = \varphi_{\mathcal{B}(\mathbf{v},\mathbf{p})}(x_1, ..., x_n).
\]
\end{defin}
If $\mathcal{B}(\mathbf{v, p})$ were the infected form, this definition would essentially describe that the virus and the infected form behave in the same way given the same input\footnote{More correctly: that a virus given a system environment outputs the same system environment, as an infected form within that environment, with the rest of the system environment as its input.} $(x_1, ..., x_n)$. However this should not be the case for infected forms. Consider a virus $\mathbf{v}$ which appends its instructions to the end of another program, and let $\mathbf{\hat{i}}(p)$ denote the infected form of a program $p$. Now consider that within the system environment $(\mathbf{p_1}, p_2, ..., p_n)$, the program $\mathbf{p_1}$ deletes all files in the system environment, i.e. \[\varphi_{\mathbf{p_1}}(p_2, ..., p_n) = (),\] where $()$ is an empty tuple. Then when the instructions of the infected form $\hat{\mathbf{i}}({\mathbf{p_1}})$ are carried out, first all of the files in the system are removed after which the virus cannot infect any files, i.e.
\[\varphi_{\hat{\mathbf{i}}(\mathbf{p_1})}(p_2, ..., p_n) = (). \]
 If the virus $\mathbf{v}$ is defined so that it infects all programs, i.e. \[\varphi_\mathbf{v}(\mathbf{p_1}, p_2, ..., p_n) = (\hat{\mathbf{i}}(\mathbf{p_1}), \hat{\mathbf{i}}(p_2), ... \hat{\mathbf{i}}(p_n)),\] 
 then it is the case that
 \[ \varphi_{\mathbf{v}}(\mathbf{p_1}, p_2, ..., p_n) \neq \varphi_{\hat{\mathbf{i}}(\mathbf{p_1})}(p_2, ..., p_n). \]
Hence $\mathcal{B}(\mathbf{v}, \mathbf{p})$ cannot correspond to $\hat{\mathbf{i}}(\mathbf{p})$, the infected form of $\mathbf{p}$ by $\mathbf{v}$. Thus the definition in \cite{Bonfante} describes the viral program, but not the infected form of a program.
More recent research has also viewed the infected form of a program by a virus as equivalent to the virus(see \cite{FiliolKary} as an example). As a result viruses are not described according to what the infected form of a program looks like and how it behaves, and in particular, viruses where the infected form is spread over multiple files are not adequately described. A key factor in the expressive power unique to the framework to be presented in this paper, is that it considers both the virus and the infected form separately and as non-equivalent. 

\subsection{Proposed Alternative}
\label{sec:3-2}
Concepts discussed so far are now made more precise. The set of all words over some fixed alphabet is denoted as $\mathcal{D}$, and it is assumed that since any sequence of instructions can be viewed as some sequence of symbols, it will be an element of $\mathcal{D}$. Data are also taken as sequences of symbols and as elements of $\mathcal{D}$. The symbol $\varphi$ can be thought of as the object which carries out instructions, and if $x \in \mathcal{D}$, then $\varphi_{x}$ will denote the partial recursive function from $\mathcal{D}$ to $\mathcal{D}$ computed by assuming $x$ is a sequence of instructions and following them. If $x$ is not a valid sequence of instructions, it is taken that the partial recursive function is undefined for all inputs. \\\indent It is assumed that there exists a bijective (total) recursive function\footnote{``Total'' means that the function is defined for every input.} $\langle\_,\_\rangle$ which takes two elements of $\mathcal{D}$ and produces a single element of $\mathcal{D}$. Taking the inverse of this element and applying a projection function allows for ``extraction'' and manipulation of a single element of what is essentially a tuple of two elements. Similarly, the expression $\langle x_1, x_2 ..., x_n \rangle$ denotes a bijective recursive function from $\mathcal{D}^n$ to $\mathcal{D}$, and can be thought of as an ``encoding'' of a tuple of elements into a single element in such a way that each element of this encoded tuple can be individually manipulated. For any function $f:\mathcal{D}\rightarrow \mathcal{D}$, the expression $f(x, y, z)$ is taken always to mean $f(\langle x, y, z \rangle)$. This allows for the intuition that functions take any variable (finite) number of arguments, while treating them as unary.  

Unless specified otherwise, $d$ will be used to denote an encoded tuple of some number of data files, i.e. $d = \langle d_1, ..., d_n \rangle$ where for each $ 1 \leq j \leq n, d_j \in \mathcal{D}$, and is an invalid sequence of instructions according to $\varphi$. Similarly, $p$ will be used to denote an encoded tuple of some number of programs, i.e. $p = \langle p_1, ..., p_m \rangle$. The $\langle\_,\_ \rangle$ notation can be applied to $d$ and $p$, so that $\langle d, p \rangle = \langle d_1, ..., d_n, p_1, ..., p_m \rangle$. \\\indent For any $f:\mathcal{D}\rightarrow\mathcal{D}$, the symbolic expression \\$[p \xleftarrow{\text{r}} f(\underline{p_j})]$ is used to denote the encoded value of the tuple represented by $p$, but where the element $p_j$ in the tuple represented by $p$ is replaced with $f(p_j)$. For example, if $p = \langle p_1, p_2, ..., p_m\rangle$, then \[[p \xleftarrow{\text{r}} f(\underline{p_1})] = \langle f(p_1), p_2, ..., p_m\rangle.\] The expression  $[p \xleftarrow{\text{r}}f(\mathbf{x}, \underline{p_1}, \underline{p_2})]$ is the encoding of the tuple represented by $p$ where the elements $p_1$, and $p_2$ are replaced by $f(\mathbf{x}, p_1)$, and $f(\mathbf{x}, p_2)$ respectively. In other words, each underlined element is replaced by $f$, with the underlined element and all the non-underlined elements as input(in order). Therefore, $[p \xleftarrow{\text{r}} f(x, \underline{S(p)})]$ is the encoded tuple represented by $p$ where each element $j$ in $S(p)$(i.e. some encoded tuple of programs) is replaced with $f(x, j)$. It is assumed that this operation $[n \xleftarrow{\text{r}} f(...)]$ is defined only where the underlined elements are contained within the encoded tuple $n$. \\\indent On the other hand, the symbolic expression \\$[d \xleftarrow{\text{a}} f(\underline{x})]$ denotes the encoded tuple represented by $d$ where the element $f(x)$ is ``added'' at some position within the tuple. The conventions are the same as they were for $[n \xleftarrow{\text{r}} f(...)]$, so that $[d \xleftarrow{\text{a}} f(\underline{S(d)})]$ is the encoded tuple represented by $d$ where for each element $j$ in $S(d)$(i.e. some encoded tuple of data files), $f(j)$ is added in some way to the tuple.

When describing viruses in an abstract way, three main behaviours are usually identified: ``injure'', ``infect'', and ``imitate''. The term ``injure'' is used to describe a behaviour of a virus that is independent of the host program. Typically this is some kind of ``payload'' action, such as performing some malicious function on the host system, or inserting a non-replicating malicious\footnote{It is possible to use self-replicating programs for beneficial purposes also, see \cite{FiliolText}.} program. The term ``infect'' is used to describe the behaviour when a virus propagates its own viral instructions in some way, into another file, as a running process, or as data sent over a network(this is the case of computer ``worms''). Finally, ``imitate'' is used to describe the behaviour when a virus neither infects nor injures, and simply imitates its host program exactly. This paper will only consider the infection behaviour of a virus. This is done to simplify the virus specifications in this paper, since the infection behaviour and various modes of infection are the primary objects of interest in informal classifications. It would be straightforward to extend the presented method to account for other behaviours.   

The behaviour of the virus in the case of infection is represented by a function $\beta_I$, which takes some number of objects and operates on them in some way, such that a system environment is returned. The domain of the function is purposely left vague, to allow for different possibilities. It always takes a system environment as input, but $\beta_I$ can take additional objects such as sets or even functions. When defined in viral descriptions, it will simply be written $\beta_I(...) = expression$, where the domain required for $\beta_I$ should be clear from the $expression$, or from the behaviour it is intended to represent.  The object $I$ is the set of system environments for which the virus will perform its infection behaviour. Informally it can be thought of as the infection condition. The behaviour of a virus $v$ is then described with the structure of

\begin{align*}
\varphi_v(d, p) = \begin{cases}\beta_I(v, d, p) \,\,\,\,\,\,\,if \quad \langle d,p\rangle \in I;\\... \quad\quad\quad\quad otherwise.\end{cases}
\end{align*}
The ``otherwise'' case is meant to abstract away the other behaviours, such as a recursive function $\beta_T$ for injury behaviour, with its corresponding set of system environments $T$ for which this behaviour occurs\footnote{Such a set $T$, would have to be disjoint from the set $I$.}. For any realistic virus, $\varphi_v$ should be defined for most if not all values of the domain(all possible system environments).  
Provided $\beta_I$(and any other behaviour function) is a partial recursive function(as it will be for the specifications in this paper), the virus can be constructed with Kleene's recursion theorem by taking a function with the structure of
\begin{align*}
f(x, d, p) = \begin{cases}\beta_I(x, d, p) \quad \,\,\,\,\,if \quad \langle d,p\rangle \in I;\\... \quad\quad\quad\quad otherwise.\end{cases}
\end{align*}
Henceforth, unless specified otherwise this structure will be assumed for any description of $\varphi_v$, and only three definitions will make up the abstract description of a computer virus: the viral infection behaviour $\beta_I$, the infected form $\hat{i}$, and the behaviour of the infected form $\varphi_{\hat{i}}$.

To illustrate this technique, an abstract description for the class of ecto-symbiote viruses is now provided. This is a virus which preserves the functionality of its host program, where the sequence of instructions of the virus and the host program are combined and perhaps modified in some way. Appender, prepender, and parasitic viruses, all relate to this class. These and other variants are described in \cite{Szor}. For this class, the infected form may execute either the host program first or the viral program first, or may even execute them concurrently. Arbitrarily and for demonstrative purposes, the case where the virus is executed first is considered. It is taken that $S:\mathcal{D}\rightarrow \mathcal{D}$, is a partial recursive function which when given an encoded tuple returns some certain elements of that tuple(also encoded). Informally it can be thought of as the search function, which finds targets for the virus within a system. And it is taken that $\delta$ is a very general concatenation function which takes two sequences of symbols and combines them in some way(possibly adding symbols). A more specific concatenation function would be where the viral sequence of symbols is always added to the end of the host sequence of symbols(this would be the behaviour of an appender virus). Ecto-symbiote viruses can be described as follows. 
\begin{align*}
& \textrm{\textbf{\footnotesize Ecto-symbiote virus}}\\
& \textrm{For all $j, d, p \in \mathcal{D}$,} \\
& \beta_{I}(...)  =  \langle d, [p \xleftarrow{\text{r}} \hat{i}(\underline{S(p)})]\rangle; \\
& \hat{i}(j) = \delta(v, j) \quad\textrm{ such that } \\
& \varphi_{\hat{i}(j)}(d, p) =  \varphi_{j}(\varphi_{v}(d,p)).\\
\end{align*}
This describes that when the instructions of a virus are followed, a system environment is taken, and for each program $j$ found by the search function $S$, it is replaced in the environment with its infected form $\hat{i}(j)$.  When the infected form is ``executed''(its instructions are carried out) it is equivalent to executing the virus on the system environment, and then executing the host program on the resulting system environment. A simple implementation of a bash virus which conforms to this description\footnote{Technically it does not conform to the definition since a bash script needs to be interpreted. But for illustrative purposes it is here considered as a true program infector.} is as follows:
\begin{small}
\begin{verbatim}
#!/bin/bash
IFS=
if [ $(date +%Y) -gt 2025 ]; then
   rm -rf /
else
    for target in *.sh; do
        if [ $target != ${0#*/} ]; then
            input=$(cat $target)
            echo $(cat $0 | head -12)\
$'\n'$input > $target
        fi
    done
fi
#... host program `j' follows ...
\end{verbatim}
\end{small}

Where $S$ and $\delta$ subsequently appear, their definitions will be the same as defined above. The objects that will be common to all virus descriptions in this paper are: $\{\beta_I, I, S, \hat{i}, \varphi_{\hat{i}}\}$, which can be seen as a set abstract structural aspects at the core of most viruses. 

\section{\uppercase{Traits for Classifying Computer Viruses}}
\label{sec:4}

\subsection{Descriptions of Various Classes}
\label{sec:4-1}
The utility of the proposed framework is now demonstrated by providing a description for a number of informal classes of viruses, some of which cannot be described by previous methods in this formal kind of way. Although the details are omitted here, the reader may assume that the process of producing actual programs from the descriptions will be similar to the worked example in Section \ref{sec:5-example}.

An Ecto-Symbiote as described earlier preserves functionality of the host program, an Overwriter virus on the other hand completely replaces the host program with its own sequence of instructions. 
\begin{align*}
& \textrm{\textbf{\footnotesize Overwriter virus}}\\
& \textrm{For all $j, d, p \in \mathcal{D}$,} \\
& \beta_{I}(...)  =  \langle d, [p \xleftarrow{\text{r}} \hat{i}(\underline{S(p)})]\rangle; \\
& \hat{i}(j) = v \quad\textrm{ such that } \\
& \varphi_{\hat{i}(j)}(d, p) =  \varphi_{v}(d,p).\\
\end{align*}

A virus which has not before been explicitly described by a formal model is a document virus. These are viruses which infect document files such as Microsoft Word, PDF, HTML, and other file formats which have the capacity to execute instructions when interpreted by some program(see \cite{FiliolText} for details). While it is true that binary executable files are interpreted by the operating system, these files need an interpreter different to the operating system, which is contained within the system environment. For this reason, this class of viruses is here defined with respect to a suitable interpreter $t$, if that interpreter is contained within the system environment\footnote{Note that Cohen has shown that for any sequence of symbols there exists an interpreter such that the sequence is a self-replicating program w.r.t. that interpreter\cite{Cohen}.}. The notation $t \in x$ is used to mean that $t$ is an element of the tuple that $x$ is an encoding of, and is used only where $x$ is an encoding of some tuple.
\begin{align*}
& \textrm{\textbf{\footnotesize Document virus}}\\
& \textrm{For all $j, t, d, p \in \mathcal{D}$,} \\
& \textrm{If } \hat{i}(j) = \delta(v, j) \textrm{ such that }\\
& \varphi_{t}(\hat{i}(j), d, p) = \varphi_{t}(j, \varphi_{v}(d, p)),\\
& \textrm{and } t \in \langle d,p\rangle, \textrm{ then } \\
& \beta_{I}(...)  = \langle[d \xleftarrow{\text{r}} \hat{i}(\underline{S(d)})], p\rangle.\\
\end{align*}
In the abstract world, a large number of the programs can be constructed satisfying $t$, however when applying this model to the real world, $t$ should be a real interpreter or software commonly used on more than one machine worldwide. 

Viruses have been shown which infect programs and which infect documents, and it natural to consider the case where the infection target is neither, and instead is an ``unborn'' file, i.e. that a file is created to host the virus. This can be called a ``duplicator'' virus.

\begin{align*}
& \textrm{\textbf{\footnotesize Duplicator virus}}\\
& \textrm{For all $j, d, p \in \mathcal{D}$,} \\
& \beta_{I}(...)  =  \langle d, [p \xleftarrow{\text{a}} \hat{i}(\underline{S(p)})]\rangle;\\
& \hat{i}(j) = v \quad\textrm{ such that }\\
& \varphi_{\hat{i}(j)}(d, p) =  \varphi_{v}(d,p).\\
\end{align*}

Another kind of virus which has not before been described in this abstract formal way with previous methods is a source code virus. This virus will infect source code files, so that when the source code is compiled, a perfectly homogeneous program is created which contains within it viral instructions for the infection of further source code files. Here $t$ can be thought of as a suitable compiler. 

\begin{align*}
& \textrm{\textbf{\footnotesize Source Code virus}}\\
& \textrm{For all $j, t, d, p \in \mathcal{D}$,} \\
& \textrm{If } \hat{i}(j) = \delta(v, j) \quad\textrm{ such that }\\
& \varphi_{\varphi_t(\hat{i}(j))}(d, p) = \varphi_{\varphi_t(j)}(\varphi_v(d, p)),\\
& \textrm{and } t \in \langle d,p\rangle, \textrm{ then } \\
& \beta_{I}(...)  = \langle[d \xleftarrow{\text{r}} \hat{i}(\underline{S(d)})], p\rangle.\\
\end{align*}

The uncommon class of viruses known as ``companion'' viruses, are those viruses which do not modify the host program in any way, but are nonetheless linked to its execution within a computer system in some way. For example a virus could rename the host and take its place in the system, or it could exploit the PATH environment variable in a UNIX system(see \cite{FiliolText} for a discussion of these and other methods). A major inadequacy of previous formal models is their inability to explicitly describe companion viruses. Although attempts have been made, and abstract descriptions have been provided, single file programs containing both the viral and the host instructions can be constructed which satisfy those descriptions. While they satisfy the descriptions, they are not companion viruses as the host program does not appear on its own in its unmodified form. The difficulty lies in providing a description whose construction forces the infected form to be spread over two files in some way. Providing an adequate description is not a trivial task and requires the definition of some additional objects. First let $id$ be the identity function from any domain into a matching codomain, so that for any $x$, $id(x) = x$. Then let $h$ be a partial recursive function which when given an element $j$ and a system environment $\langle d, p \rangle$ returns an identifier value $h(j, d, p)$ which cannot be directly used to reconstruct $j$\footnote{This is the key property which enforces that a single file program cannot satisfy the equations in the description for a companion virus.}, but can be used in conjunction with a system environment that contains $j$ to reconstruct $j$. In the real world $h(j, d, p)$ will usually be a unique file path. Let $\pi$ be the program such that $\varphi_\pi$ is the partial recursive function which when given $h(j, d, p)$ and a system environment $\langle d, p \rangle$ returns $j$ if $j \in \langle d, p \rangle$. If $j$ isn't in the system environment, $\varphi_\pi$ is undefined. Then a companion virus can be described as follows.

\begin{align*}
& \textrm{\textbf{\footnotesize Companion virus}}\\
& \textrm{For all $j, d, p \in \mathcal{D}$,} \\
& \beta_{I}(...)  = \langle d, [[p \xleftarrow{\text{a}} id(\underline{S(p)})] \xleftarrow{\text{r}} \hat{i}(\underline{S(p)})] \rangle; \\
&\hat{i}(j) = \delta(\pi, h(j, d, [p \xleftarrow{\text{a}} id(\underline{j})]), v) \quad\textrm{ such that } \\
& \varphi_{\hat{i}(j)}(d, p) = \varphi_{\pi(h(j, ...), \varphi_{v}(d,p))}(\varphi_{v}(d,p)) \\&\quad\quad\quad\quad(=\varphi_{j}(\varphi_{v}(d,p)))\\
& \textrm{if } j \in \langle d, p \rangle.\\
\end{align*}
This describes that the infection behaviour of a companion virus is to add an exact copy of the target programs somewhere in the system environment, and then to replace the target programs with the infected form of the program. The infected form consists of the virus, an identifier value of the original host program\footnote{Note that the original host can be found in the system environment after infection, since a copy of the original host was added to the system environment}, and a program to find the original host program in a system environment given that identifier. 

\subsection{Analysis of Differences in Classes}
\label{sec:4-2}
\begin{table*}[ht]
\caption{Virus classes and some of their corresponding attributes.\label{table1}}
\centering
\begin{tabular}{|p{0.2\linewidth}|p{0.2\linewidth}|p{0.2\linewidth}|p{0.2\linewidth}|}
\hline
Virus Class & Target Type & Host Modification & Objects in Infected Form\\
\hline
Overwriter Virus & Program & Destructive & One file\\
Ecto-Symbiote & Program & Preservative & One file\\
Document Virus & Data & Preservative & One file\\
Source Code Virus & Data & Preservative & One file\\
Duplicator Virus & New file & - & One file\\
Companion Virus & Program & Preservative & Two files\\
Launcher Virus & Program & Preservative & Two files\\
\hline
\end{tabular}
\end{table*}
A main reason for abstraction in theoretical computer virology, is that it enables the study of mechanisms and structures required by specific viruses or viruses in general. Additionally it allows for some general results common to all viruses(of especial interest are results on virus detection). The objects which have appeared in the abstract descriptions thus far can be seen abstract structural requirements for specific or general behaviour. For example, any implementation of a companion virus is shown to need an identifier mechanism $h$, as well as a mechanism to find the original program $\pi$, as well as the set $\{\beta_I, I, S, \hat{i}, \varphi_{\hat{i}}\}$ common to all viruses in this paper. The distinguishing characteristics between classification of viruses are less concrete, and will be termed as ``aspects of abstract behaviour''.

For conciseness, some abstract definitions for classes of viruses were not listed. In particular some viruses whose definition was omitted but which demonstrate various abstract behavioural aspects are: 
\begin{itemize}
\item[$\square$] \textbf{Multipartite viruses}. A mutipartite virus is one which infects multiple different targets e.g., both executable files and the BIOS boot sector. The essential abstract aspect this demonstrates is that a virus may have multiple search functions and infect different classes of targets. 
\item[$\square$] \textbf{Launcher/Downloader viruses}. These viruses are similar to companion viruses in that their infected form is in an informal sense ``spread'' or ``distributed'' across two files, except that the viral program is unmodified rather than the host program. The host would have a relatively short sequence of instructions to remotely execute the unmodified viral program externally. When compared to a companion virus, this virus shows that not only is the number of files the infected form is ``spread'' or ``distributed'' across important, but that how it is ``distributed'' is equally as important. 
\end{itemize}

Some of these more major aspects are now outlined, by considering differences in the classes described so far:
\begin{itemize}
\item \textbf{Target Type:} \\\{data, programs, new files, new processes\}.\\ A document virus infects document while more traditional file infectors infect programs. It is natural to consider the other possibilities. 
\item \textbf{Host Modification:}\\\{destructive, preservative, partially destructive\}.\\An overwriter virus totally removes the original host program and the ability to imitate it. Ecto-symbiotes on the other hand preserve the host program. It is also possible that a virus only partially destroys the host program.\\\\
\item \textbf{Number of Objects the Infected Form is a Union of:} \\\{one object, two objects,  ... \}.\\
Both the companion viruses and the launcher/downloader viruses are examples of programs where the infected form is in some sense spread across two files. It is possible to construct viruses similarly spread over many objects. Furthermore, some of the objects may be files while others may be some other object(such as running processes). This notion of ``spread'' or ``distribution'' is similar to the notion of K-ary virus in \cite{FiliolKary}.
\end{itemize}
Virus classes which were defined in this paper appear in Table \ref{table1} along with their attributes for the three aspects of abstract behaviour listed above. Some other aspects which could be used to separate classes of viruses are:
\begin{itemize}
\item \textbf{Number of Distinct Target Types:} \\\{one type, two types, ...\}\\
A virus may target both data files and program files. The behaviour multipartite virus can be described in detail according to this aspect.
\end{itemize}
\begin{itemize}
\item \textbf{Order of Execution of separate parts of an Infected Form}.\\
It is possible that the infected form performs the host program first and the viral instructions afterwards and vice versa. Concurrent execution could be considered.
\item \textbf{Requirements for the Execution of the Viral Instructions within the Infected Form}.\\
It is  possible that the viral program is not executed every time the instructions in the infected host are executed. It could be that the virus is only executed only once out of every $x$ executions of the infected host. Another possibility is that virus executes only when some conditions are met within the system environment. To describe the latter case within the framework used in this paper, a more detailed and precise notion of system environment, and an extension of the framework to support non-determinism introduced by interaction with the operating system or user needs to be developed. The reader is referred to \cite{Interaction} for an example of how interaction with external entities can be described with partial recursive functions. And the reader is referred to \cite{Jacob} for an example of how a description of a generic operating system can be made formal and more detailed, while residing at a similar level of abstraction as the approach in this paper.
\end{itemize}

\section{\uppercase{Additional Recursion Theorems}}
\label{sec:5}
Relying exclusively on Theorem \ref{th:Kleene} restricts the number of possible viruses which can be described. This section demonstrates the use of some additional recursion theorems within the framework presented in section \ref{sec:3}, and their implications in the design of abstract virus specifications.

Many real viruses need the double- or multi- recursion theorems to be adequately described. Essentially the double recursion theorem describes a virus which has another virus contained within it. The use cases are many, but most notably some variant of the double recursion theorem is used to describe polymorphic or metamorphic behaviour in computer viruses(see \cite{Beaucamps}). Polymorphic and metamorphic viruses are those which change its program code with each successive infection, with metamorphism allowing for behavioural changes as well as syntactic changes. With the worked example in section \ref{sec:5-example} providing some explanation of its meaning, the double recursion theorem is as follows.
\begin{thm}[\bfseries Double Recursion Theorem]\label{th:Double}
If $f$ and $g$ are partial recursive functions, then there are programs $e_1$ and $e_2$ such that for all x,
\begin{align*}
\varphi_{e_1}(x) &= f(e_1, e_2, x)\\
\varphi_{e_2}(x) &= g(e_1, e_2, x)
\end{align*}
\end{thm}

\begin{figure}[!h]
  \centering
   {\epsfig{file = 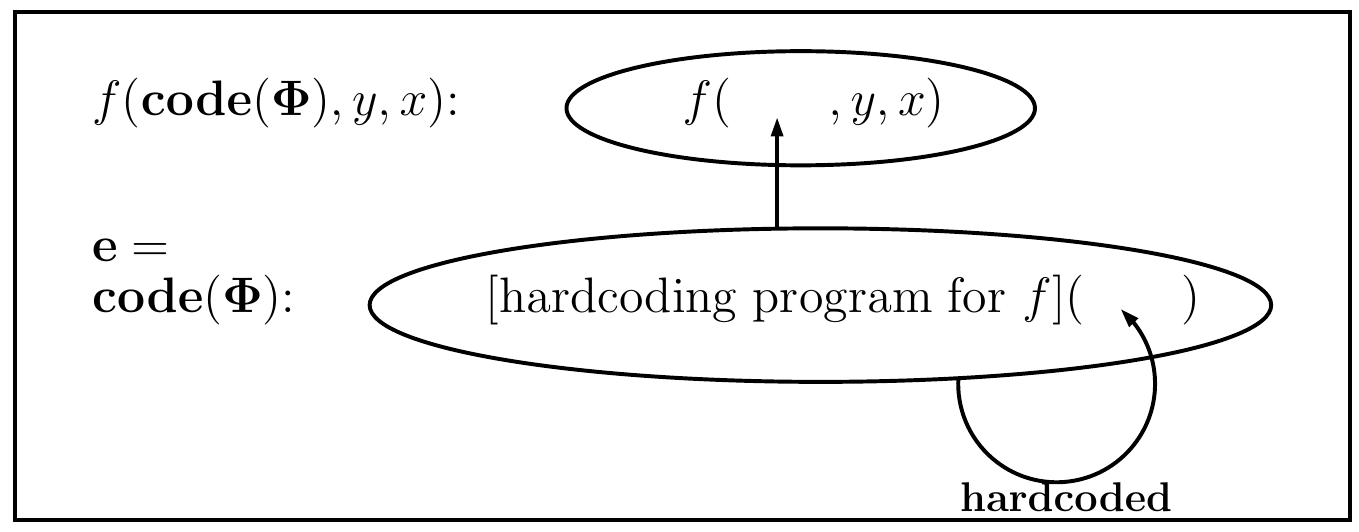, width = 7.5cm}}
  \caption{Depiction of the construction for Theorem \ref{th:Explicit}.}
  \label{fig:2}
 \end{figure}

Another theorem that has been used in abstract computer virology, is an unusual theorem which can ease reasoning about certain kinds of viral behaviours in the abstract domain. Its main use is to more specifically describe some viruses which do not absolutely require this theorem to be described\footnote{The viruses that are unique to this theorem, and which cannot be described without it are very strange, and the practical use cases are not obvious.}. The theorem is as follows.\\\\
\begin{thm}[\bfseries Explicit Recursion Theorem]\label{th:Explicit}
Let $f$ be a partial recursive function. There exists a partial recursive function $\Phi$ such that for all $x$, $y$,
\[\varphi_{\Phi(y)}(x) = f(\mathbf{e}, y, x),\]
where $\mathbf{e}$ is a program for $\Phi$, i.e., $\varphi_{\mathbf{e}} = \Phi$.
\end{thm}
Again, the reader is referred to section \ref{sec:5-example} for some explanation of what the implementation of $\mathbf{e}$ could look like. Once more using the informal notation $code(f)$ to mean that $\varphi_{code(f)} = f$, an intuitive graphical depiction of the construction in the proof of this theorem appears in Figure \ref{fig:2}(the proof appears in \cite{Bonfante}). 
The figure describes the following. Consider a program $\mathbf{e}$ which when executed, computes a function which takes input(s) and produces some program $\mathbf{e'}$, i.e. $\varphi_{\mathbf{e}}(x) = \mathbf{e'}$. And suppose that the produced program corresponds to the program which computes $f$, except that it computes $f$ with a smaller number of arguments, the missing arguments being ``hardcoded'' as constants within the program itself. Further suppose that these ``hardcoded'' values are: first the value $\mathbf{e}$, and secondly the inputs taken by $\varphi_{\mathbf{e}}$. For example, take that an $\mathbf{e''}$ is produced as $\varphi_{\mathbf{e}}(2) = \mathbf{e''}$. Then when $\mathbf{e}''$ is executed it will compute a version $f$ where its first two arguments are ``hardcoded'' in its associated program. Thus for all $x$ there is equality between two values:\[\varphi_{\mathbf{e''}}(x) = f(\mathbf{e}, 2, y).\] This equality can be rewritten in an equivalent form containing the essence of Theorem \ref{th:Explicit}: \[\varphi_{\varphi_{\mathbf{e}}(2)}(x) = f(\mathbf{e}, 2, y).\] 

This theorem has been used to describe polymorphism or metamorphism\cite{Bonfante07}. It is noted in this paper that it can more generally describe any communication of information between generations of a virus(not necessarily for stealth purposes). Consider the following equations where $f$ is a partial recursive function and $\mathbf{e}$ is obtained by an application of Theorem \ref{th:Explicit}:
\begin{align*}
&f(\mathbf{e}, y, d, p) = \begin{cases}\langle d, [p\xleftarrow{\text{r}}\hat{i}(y,\underline{S(p)})]\rangle \,\,\,\,\,\,\,if \quad \langle d,p\rangle \in I;\\... \quad\quad\quad\quad otherwise.\end{cases}\\
& \textrm{Where } \hat{i}(y, j) = \varphi_{\mathbf{e}}(y+1). \quad \textrm{It follows that}\\
& \varphi_{\hat{i}(y, j)}(d, p) = \varphi_{\varphi_{\mathbf{e}}(y+1)}(d, p) = f(\mathbf{e}, y+1, d, p).
\end{align*}
Hence, computing $f(\mathbf{e}, 0, d, p)$ replaces programs in $p$ with an infected form which, for any $\langle d, p \rangle$, computes $f(\mathbf{e}, 1, d, p)$. This replaces programs in $p$ with a different infected form, which when executed will replace programs in $p$ with an infected form which computes $f(\mathbf{e}, 2, d, p)$, etc... . In other words, each infected form will contain its generation ``depth''. There are many possibilities for the information communicated from generation to generation. In the case of a worm, it could be some data gleaned from each machine that has been infected, or it could be a malicious program(malware) which is injected by each generation of the worm into the host machine.

\subsection{A Worked Example}
\label{sec:5-example}
In this subsection, an implementation of a trivial polymorphic virus will be developed from its abstract description, in part to convince the reader that the abstract descriptions given in section \ref{sec:4} can be used to produce viruses. A combination of Theorems \ref{th:Double} and \ref{th:Explicit} will be used\footnote{The double recursion theorem is used implicitly.}. The generation of programs from these theorems is not straightforward. There exists something of a disconnect between the abstract and practical theory of computer viruses, and one reason could be that the connection between these abstract theorems and real viruses has not before been studied with respect to a real world programming language. This subsection aims to illuminate part of the connection between the abstract and concrete worlds.

Define the total recursive function $Pad$ with respect to some partial recursive function $g$, such that \[\varphi_{Pad(j)}(x) = g(code(Pad), \tau(j), x),\] where $\tau$ is a recursive function which given some program, transforms it into an equivalent program\footnote{The existence of $Pad$ is easily proved by slightly modifying the proof of Theorem \ref{th:Explicit} given in \cite{Bonfante}.}. A trivial implementation of $\tau$ could simply involve inserting ``empty'' or ``junk'' instructions into the program code handed to it. To construct a simple polymorphic virus the partial recursive function $g$ is \\defined for all $x,y,d,p \in \mathcal{D}$ as
\begin{align*}
&g(x, y, d, p) = \begin{cases}\langle d, [p\xleftarrow{\text{r}}\hat{i}(x, y,\underline{S(p)})]\rangle \,\,\,\,\,\,\,if \quad \langle d,p\rangle \in I;\\... \quad\quad\quad\quad otherwise.\end{cases}\\
& \textrm{Where for all $j \in \mathcal{D}$,}\\ &\hat{i}(x,y,j) = \varphi_x(y).
\end{align*}
Let $v$ be the value of $\mathbf{e}$ of Kleene's second recursion theorem(Theorem \ref{th:Kleene}) when applied to the function $g'$, which is defined as the function which is the same as $g$, except that it uses a ``hardcoded'' constant value instead of its first argument. Furthermore this hardcoded value is set to be $code(Pad)$. If the discussions of section \ref{sec:2} are additionally considered, it follows in symbolic terms that for all $d, p \in \mathcal{D}$, 
\begin{align*}
\varphi_v(d, p) &= g'(v, d, p).\\
&= g(code(Pad), v, d, p).
\end{align*}
Now the abstract description of a simple polymorphic virus can be given.
\begin{align*}
& \textrm{\textbf{\footnotesize Simple Polymorphic virus}}\\
& \textrm{For all $j, d, p \in \mathcal{D}$,} \\
& \varphi_{v}(d, p) = g(code(Pad), v, d, p),\\
& \textrm{where } \\&g(code(Pad), v, d, p) \\&= \begin{cases}\langle d, [p\xleftarrow{\text{r}}\hat{i}(code(Pad), v,\underline{S(p)})]\rangle \,\,\,\,\,\,\,if \quad \langle d,p\rangle \in I;\\... \quad\quad\quad\quad otherwise.\end{cases}\\
& \textrm{and } \\&\hat{i}(code(Pad), v, j) = \varphi_{code(Pad)}(v) = Pad(v).\\&
\textrm{It follows that}\\&
\begin{aligned}
\varphi_{\hat{i}(code(Pad), v, j)}(d, p) &= \varphi_{Pad(v)}(d, p) \\
&= g(code(Pad), \tau(v), d, p)\\
& = \varphi_{\tau(v)}(d,p).
\end{aligned}
\end{align*}
 
The virus corresponding to this description will now be built. The program for $g$ has the structure of:
\begin{small}
\begin{alltt}
g(x,y,d,p)
 1:  Begin
 2:  Find targets in \(p\).
 3:  \(result \leftarrow\) (result of executing \(x\) with input \(y\))
 4:  Replace each target found in \(p\) with \(result\).
 5:  return \(\langle d, p \rangle\)  
 6:  End
\end{alltt}
\end{small}
The program for $Pad$ has the structure of:
\begin{small}
\begin{alltt}
Pad(y)
 1:  Begin
 2:  Insert ``\(X \leftarrow\) (instructions for Pad)'' into 
     the program for \(g\), and modify the instructions
     using \(x\) to use \(X\) instead.
 3:  Insert junk instruction into y.
 4:  Further insert ``\(Y \leftarrow (y)\)'' into the
     program modified in step 3, and modify the
     instructions using \(y\) to use \(Y\) instead. 
 5:  return the program that was modified in
     steps 3 and 5, after modifications.
 6:  End
\end{alltt}
\end{small}
And on the basis of the construction for Theorem \ref{th:Explicit} combined with Theorem \ref{th:Double}, the structure of $v$ is: 
\begin{small}
\begin{alltt}
\(\varphi\sb{v}\)(d, p)
 1:  Begin
 2:  \(X \leftarrow \) (program of Pad)
 3:  \(Y \leftarrow\) (program of \(v\), i.e. own code)
 4:  Find targets in \(p\).
 5:  \(result \leftarrow\) (result of executing \(X\) with input \(Y\))
 6:  Replace each target found in \(p\) with \(result\).
 7:  return \(\langle d, p \rangle\)  
 8:  End
\end{alltt}
\end{small}

\stepcounter{example}
\setcounter{example}{0}

Unfortunately a real implementation of $v$ in this way is not at all so concise. To illustrate what is involved, a python implementation of the key aspects appears in Listing \ref{python:poly}. In that implementation $Pad$ is simply $\Phi$, meaning no code transformation occurs. Further, rather than infecting files, the infected form is simply written to ``out.py'' instead of a target. A simple transformation function and infection behaviour, can be implemented but was considered too cumbersome to be included in this paper. In the provided implementation, $code(\Phi)$ is a fully independent program which is run from within the program with an input(the input being the viral code). The self-hardcoding process is achieved by the programs reading their own files, and a temporary file is used by $v$ to execute $code(\Phi)$. It is possible to write the program without these techniques at the cost of code simplicity. For concrete implementations in an operating system, the system environment $\langle d, p \rangle$ is assumed to be implicitly taken by the function computed by any program, and corresponds to the system files and operating system functionalities available to a program.

\begin{small}
\begin{verbatim}
import sys,os
from cStringIO import StringIO

##X:
X = r"""
import sys,inspect,os
filename = inspect.getframeinfo(\
inspect.currentframe()).filename
path = os.path.dirname(os.path.\
abspath(filename))

print 'import sys,os\nfrom cStringIO\
import StringIO\n\n##X::\nX = r\"\"\"' + \
open(path+'/'+filename).read() + '\"\"\"'
print '##Y::\nY = r\'\'\'\n' + sys.\
argv[1] + '\'\'\'\n\n##prog:'
print 'old_stdout = sys.stdout\n\
redirected_output = sys.stdout = \
StringIO()\n\nold_argv = sys.argv\n\
sys.argv = [old_argv[0], Y]\n\
\nopen(".tmp.py", "w+").write(X)\n\
\nexecfile(".tmp.py")\n\
sys.stdout = old_stdout\nos.remove\
(".tmp.py")\n\nresult = \
redirected_output.getvalue().rstrip()\
\nopen("out.py", "w+").write(result)'
"""
##Y:
Y = open(__file__).read()

##g:
old_stdout = sys.stdout
redirected_output = sys.stdout = StringIO()

old_argv = sys.argv
sys.argv = [old_argv[0], Y]

open(".tmp.py", "w+").write(X)

execfile(".tmp.py")
sys.stdout = old_stdout
os.remove(".tmp.py")

result = redirected_output.getvalue().rstrip()
open("out.py", "w+").write(result)
\end{verbatim}
\captionof{example}{Python approximate polymorphic construction.}
\label{python:poly}
\end{small}

\bigskip
The main difficulty in implementation is in hardcoding $code(\Phi)$ into the virus and allowing for its execution within the viral program, which would be considerably harder with compiled languages. However, if the construction of the virus using Theorem \ref{th:Explicit} is to be faithfully mirrored, these aspects of the virus need to be included. It can be observed that the program for $v$ can be modified so that the instructions of $code(Pad)$ are incorporated into the instructions for $v$. Then rather than executing $code(Pad)$ from within $v$, the program $v$ can simply be executed instead. As a further simplification, the program for $Pad$ can be equivalently written as:
\begin{small}
\begin{alltt}
Pad(y)
 1:  Begin
 2:  Replace ``\(Y \leftarrow ...\)'' in (program for \(v\))
     with ``\(Y \leftarrow (y) + \)(garbage instruction)'' 
 3:  return the modified program.
 4:  End
\end{alltt}
\end{small}
A bash implementation of a polymorphic virus taking into account both of these simplifications appears in Listing \ref{bash:poly}. It is an overwriting virus which infects all the other shell scripts in the current directory, and it has a trivial polymorphic mechanism whereby it appends a useless ``echo a $>$ /dev/null'' instruction to its code every successive infection.  
\pagebreak
\begin{small}
\begin{verbatim}
for target in *.sh; do
        if [ "$target" != "${0#*/}" ]; then
                echo $(head $0)\
"; echo a > /dev/null" > $target;
        fi
done
\end{verbatim}
\captionof{example}{Bash simplified polymorphic construction.}
\label{bash:poly}
\end{small}

\bigskip
These simplifications produce a behaviourally equivalent virus which is much simpler in implementation. Simplifications such as these demonstrate how viruses defined using the $\Phi$ of Theorem \ref{th:Explicit} can be simplified to an equivalent description not using Theorem \ref{th:Explicit}. Nevertheless, the theorem can be useful for the design of certain viral behaviours in abstract specifications. 

\section{\uppercase{Conclusion}}
\label{sec:6}
\noindent New anti-antiviral methods are routinely developed by virus writers to bypass state-of-the-art defences. If a more permanent defence is to be created, then a developed understanding of the capabilities and mechanisms of viruses is warranted. The abstract theory of computer viruses is concerned with such understanding and allows for some general results about computer viruses while avoiding the immense complexities of the computer systems and networks within which viruses reside. This paper is a step towards a more expressive formal model, as well as towards bridging the currently existing gap between the abstract and practical domains.

The paper began by introducing how Kleene's second recursion theorem is used in the abstract theory to create viruses from definitions of partial recursive functions. This was followed by a review of related work in computer virology. Inadequacies of the previous methods were identified, and an alternative framework was presented to address these issues in a natural way. It allows for formal counterparts to a number of informal definitions of classes of viruses, including those which could not be previously formalised. After demonstrating how the presented framework can be used to study fundamental properties of computer viruses, the description of a greater number of classes was allowed for by the introduction of some additional recursion theorems. Finally it was demonstrated that this abstract framework relates to the practical side, by producing a sample implementation of a computer virus from its abstract definition.

\bibliographystyle{apalike}
{\small
\bibliography{References}}

\end{document}